\documentclass[conference]{IEEEtran}
\IEEEoverridecommandlockouts
\usepackage{cite}
\usepackage{amsmath,amssymb,amsfonts}
\usepackage{algorithmic}
\usepackage{graphicx}
\usepackage{booktabs}
\usepackage{subcaption}
\usepackage{colortbl}
\usepackage{textcomp}
\usepackage{cleveref}
\usepackage{listings}
\usepackage{tcolorbox}
\usepackage{xcolor}
\usepackage{makecell}
\usepackage{xspace} 
\usepackage{tabularx}
\usepackage{multirow}
\usepackage{array}
\usepackage{url}
\usepackage{enumitem}
\Crefformat{figure}{#2Fig.~#1#3}
\Crefmultiformat{figure}{Figs.~#2#1#3}{ and~#2#1#3}{, #2#1#3}{ and~#2#1#3}

\def\BibTeX{{\rm B\kern-.05em{\sc i\kern-.025em b}\kern-.08em
    T\kern-.1667em\lower.7ex\hbox{E}\kern-.125emX}}


\definecolor{red}{rgb}{1,0,0}
\definecolor{codegreen}{rgb}{0,0.6,0}
\definecolor{codegray}{rgb}{0.5,0.5,0.5}
\definecolor{codepurple}{rgb}{0.58,0,0.82}
\definecolor{backcolour}{rgb}{0.92,0.92,0.92}
\lstdefinestyle{mystyle}{
    backgroundcolor=\color{backcolour},   
    commentstyle=\color{codegreen},
    keywordstyle=\color{magenta},
    numberstyle=\tiny\color{codegray},
    stringstyle=\color{codepurple},
    basicstyle=\ttfamily\footnotesize,
    breakatwhitespace=false,         
    breaklines=true,                 
    captionpos=b,                    
    keepspaces=true,                 
    numbersep=5pt,                  
    showspaces=false,                
    showstringspaces=false,
    showtabs=false,                  
    tabsize=2,
}

\def\name{\textsc{Metamon}\xspace}

\begin{document}

\title{\name: Finding Inconsistencies between Program Documentation and Behavior using Metamorphic LLM Queries\thanks{This work has been supported by the National Research Foundation of Korea (NRF) funded by the Korean government MSIT (RS-2023-00208998), and the Institute of Information Communications Technology Planning \& Evaluation (IITP) grant funded by the Korea government (MSIT) (RS-2022-II220995).}}

\author{\IEEEauthorblockN{Hyeonseok Lee}
\IEEEauthorblockA{\textit{School of Computing} \\
\textit{KAIST}\\
Daejeon, Republic of Korea \\
christmas@kaist.ac.kr}
\and
\IEEEauthorblockN{Gabin An}
\IEEEauthorblockA{\textit{School of Computing} \\
\textit{KAIST}\\
Daejeon, Republic of Korea \\
gabin.an@kaist.ac.kr}
\and
\IEEEauthorblockN{Shin Yoo}
\IEEEauthorblockA{\textit{School of Computing} \\
\textit{KAIST}\\
Daejeon, Republic of Korea \\
shin.yoo@kaist.ac.kr}
}

\maketitle

\begin{abstract}
Code documentation can, if written precisely, help developers better understand the code they accompany. However, unlike code, code documentation cannot be automatically verified via execution, potentially leading to inconsistencies between documentation and the actual behavior. While such inconsistencies can be harmful for the developer’s understanding of the code, checking and finding them remains a costly task due to the involvement of human engineers. This paper proposes \name, which uses an existing search-based test generation technique to capture the current program behavior in the form of test cases, and subsequently uses LLM-based code reasoning to identify the generated regression test oracles that are not consistent with the program specifications in the documentation. \name is supported in this task by metamorphic testing and self-consistency. An empirical evaluation against 9,482 pairs of code documentation and code snippets, generated using five open-source projects from Defects4J v2.0.1, shows that \name can classify the code-and-documentation inconsistencies with a precision of 0.72 and a recall of 0.48. 
\end{abstract}


\section{Introduction}

Code documentation, such as comments or docstrings, is a human-readable description of the source code and its behavior. While languages like Java and Python provide documentation formats that can be processed by machines to produce structured documents (\texttt{PyDoc} and \texttt{JavaDoc}, respectively), primarily code documentation aims to aid the human understanding of the code that they accompany~\cite{de2006documentation}. As such, it is critical that code documentation is consistent with the behavior of the code they accompany. If the documentation is out of sync with the program behavior, developers may understand the existing code incorrectly or imprecisely, potentially leading to buggy code changes. 

Finding such inconsistencies between code documentation and program behavior is an important but challenging problem, due to the fact that documentation is written in natural languages whereas program behavior is described in programming langauges. Checking the semantic consistency across natural and programming languages has been primarily reserved only for human developers. However, the cost of manual inspection means that such checks cannot be performed frequently enough, resulting in inconsistencies in real-world projects~\cite{evolutionOfCodeComments}. Existing attempts to detect these inconsistencies either depend on rule-based approaches~\cite{tan2007icomment, tan2011acomment, tan2012tcomment, ratol2017detecting} or textual similarities~\cite{corazza2018coherence, panthaplackel2021deep, rabbi2020detecting} to make the task more tractable.

Recently, LLMs have shown significant capabilities to perform logical reasoning across the barrier of natural and programming languages. In addition to successfully synthesizing code from natural language specifications~\cite{li2022competition}, LLMs can generate bug reproducing tests from bug reports written in natural language~\cite{kang2023large}, or incorporate error messages from synthesized test cases to improve the code coverage iteratively~\cite{schafer2023adaptive}. These capabilities are directly relevant to the task of checking documentation inconsistencies, and LLMs have been evaluated for the task~\cite{li2024mutation}. However, existing techniques based on LLMs tend to simply prompt LLMs to spot inconsistencies between code and documentation and falls short of actually checking the dynamic program behaviour. Given that LLMs can hallucinate~\cite{huang2023survey}, augmenting LLM-based inconsistency checking with a more concrete exploration of program behavior seems to be a missed opportunity.

This paper presents \name, an automated technique that finds inconsistencies between code documentation and program behavior via the use of a search-based test data generation technique and LLMs. To capture program behavior in more detail, instead of presenting the code verbatim in prompts, \name uses EvoSuite~\cite{Evosuite} to generate regression test cases. These test cases not only achieve high structural coverage (and thereby expose program behavior in more depth) but also capture the current program behavior in the form of assertions that record the program output. \name uses LLMs to judge whether these regression test oracles are correct or not with respect to the current documentation. To further enhance its performance, \name adopts a couple of prompt engineering strategies. First, \name uses metamorphic LLM queries: given a prompt $P$ and an answer $A$, \name will form a subsequent query prompt $P'$ and an expected answer $A'$ by transforming $(P, A)$ with Metamorphic Relations (MRs). Second, \name uses Chain-of-Thoughts~\cite{wei2022chain} and self-consistency~\cite{wang2022self} to improve the accuracy of the responses from the LLM. 

We evaluate \name using 9,482 pairs of Java methods and their documentation, taken from five open source projects in Defects4J~\cite{just2014defects4j} version 2.0.1: the dataset contains both consistent and inconsistent pairs, since we generate half of the pairs using the buggy versions. \name can classify the consistency between the code documentation and the program behavior captured in test cases with a precision of 0.72 and a recall of 0.48. An ablation study shows that both the metamorphic queries, and the advanced prompt engineering techniques, contribute positively to the performance. 

The technical contributions of this paper are as follows:

\begin{itemize}[leftmargin=10pt]
\item We present \name, an LLM-based technique that can check inconsistencies between documentation and program behavior. \name captures program behavior by generating regression test cases using EvoSuite, and uses metamorphic LLM queries and tailored prompt engineering techniques to enhance its performance.

\item We perform an empirical evaluation of \name based on 9,482 pairs of documentation and source code extracted from five projects in Defects4J v2.0.1. An ablation study shows that metamorphic LLM queries and self-consistency-based scoring all contribute to the final performance.

\item We make the implementation of \name, along with the dataset of 9,482 pairs of code documentation and method/test code snippets, publicly available for replicability: \url{https://figshare.com/s/dd17b119d40d4bf3176a}
\end{itemize}

The rest of the paper is structured as follows: \Cref{sec:background} provides an overview of the background and related research. \Cref{sec:methodology} presents our approach, \name. Detailed experimental settings and research questions for evaluating \name are covered in \Cref{sec:setup}, while \Cref{sec:results} details the findings, and \Cref{sec:conclusion} concludes.

\section{Background and Related Work} 
\label{sec:background}


\subsection{Code-and-Documentation Inconsistency}

Code documentation refers to the descriptions of specifications and requirements of the program, or explanations of the source code, written with the purpose of improving code readability, facilitating easier maintenance, supporting collaboration among developers, and enhancing code reusability. However, it is common for documentation and source code to become out-of-sync over time, which can happen when only code is updated without the documentation or vice versa~\cite{evolutionOfCodeComments}. Analyzing the consistency between documentation and source code has traditionally been a challenging task. Some approaches have used rule-based methods to detect specific types of inconsistencies, such as issues related to locking~\cite{tan2007icomment}, interrupt-related concurrency bugs~\cite{tan2011acomment}, null value-related exceptions~\cite{tan2012tcomment}, and identifier renaming~\cite{ratol2017detecting}. Other techniques have treated the inconsistency detection problem as a text similarity problem, employing machine learning models~\cite{corazza2018coherence, panthaplackel2021deep, rabbi2020detecting}, which often requires a dedicated training step.

Recently, LLMs have shown remarkable capabilities in understanding both natural and programming languages, enabling them to be applied to tasks such as code generation from specifications~\cite{li2022competition, shin2023prompt} or document generation from source code~\cite{sun2023automatic, ahmed2023improving, shin2023prompt}. This suggests that checking the consistency between code and documentation is now becoming increasingly feasible. For example, a recent study shows that GPT-4~\cite{achiam2023gpt} can identify subtle inconsistencies between code and its documentation~\cite{li2024mutation}. However, this study does not focus directly on the problem of inconsistency detection, but rather uses it as a means of assessing the code understanding capabilities of LLMs. In comparison, we propose a novel LLM-based approach that checks the consistency between program behavior (captured by regression tests) and their specifications captured in documentation, rather than directly comparing source code and documentation. We also introduce the concept of metamorphic prompting.

\subsection{Metamorphic Testing}
Metamorphic testing~\cite{segura2016survey} is a testing technique that aims to reveal faults when there is no explicit test oracle. In metamorphic testing, the correctness of a program is not based on the expected output (from oracles): rather, it is based on the relationships between different inputs and their corresponding outputs, known as Metamorphic Relations (MRs). For example, in a program that calculates the square of a number, i.e., $f(x) = x^2$, the metamorphic relation could be that the square of the negative of a number is equal to the square of the number, i.e, $a = -b \rightarrow f(a) = f(b)$. Metamorphic testing has been successfully applied to machine learning models~\cite{applis2021assessing, murphy2008properties, xie2011testing}, which are essentially untestable~\cite{Weyuker1982aa}.

In our work, we use the concept of metamorphic testing to assess the reliability of LLMs in identifying inconsistencies between program specifications and behavior. 
By examining the alignment of the LLM's responses with the expected MRs, we can assess the model's reliability and consistency in comprehending the underlying relationships between the program documentation and behavior. Note that this approach can also be seen as a form of LLM self-consistency, as the LLM should produce opposite outputs for inverted queries if it is truly consistent.


\section{methodology}\label{sec:methodology}

In this paper, we introduce \name, a novel approach that uses LLM to automatically identify inconsistencies between program documentation and behavior. Given the program documentation for a method that meets certain quality criteria (Step A), instead of directly analyzing the method's source code, \name generates regression tests to capture the current semantics of the target method (Step B). To enhance the reliability of the evaluation, \name generates two types of prompts based on metamorphic relations: the original/transformed-version prompts that ask whether the captured program behavior in the original/transformed versions of the regression test aligns well with the program documentation (Step C). Subsequently, each type of prompt is queried to LLM multiple times, and the answers are recorded (Step D). The final set of responses from the LLM is aggregated to compute a \emph{consistency} score, which numerically represents the extent to which the given specification in method documentation aligns with the captured method behavior (Step E). In the following sections, we provide more detailed explanations for each step of \name approach.

\subsection{Selecting Method Documentation}
The quantity and quality of documentation vary across projects and even down to the level of individual classes or methods, influenced by factors such as a method's complexity and its significance within the project. To ensure a fair evaluation of \name, we focus on documentation containing specifications that meet a set of minimum criteria.

Components in documentation essential for our analysis include descriptions of method input parameters and expected output values. This description is needed when constructing unit tests, which are divided into the test prefix/setup and the test oracle. The test prefix/setup initiates the method with appropriate inputs and drives the unit under test to an interesting state, while the test oracle specifies a condition that the resultant state should satisfy. These specifications are typically documented in Java using \texttt{@param} and \texttt{@return} tags. Thus, our selection process prioritizes methods whose specifications clearly delineate these aspects, ensuring \name is assessed against well-defined and actionable criteria.

\lstset{
    basicstyle=\ttfamily\scriptsize
}
\begin{figure*}[htb]
    \begin{tabular}{@{}p{0.48\linewidth} p{0.48\linewidth}@{}}
        \begin{subfigure}{\linewidth} 
\begin{lstlisting}[language=Java]
public static String getPackageName(String className) {
    if (StringUtils.isEmpty(className)) {
        return StringUtils.EMPTY;
    }

    while (className.charAt(0) == '[') {
        className = className.substring(1);
    }
    
    //
    // < ... Omitted ... >
    //
    
    if (i == -1) {
        return StringUtils.EMPTY;
    }
    // Bug! (should be substring(0, i);)
    return className.substring(1, i);     
}
\end{lstlisting} 
        \subcaption{In Lang-1f, \textbf{ClassUtils.Java}-(mutated line 306)}
        \end{subfigure}
        & 
        \begin{subfigure}{\linewidth} 
\begin{lstlisting}[language=Java]
/**
 * Gets the package name from a String.
 * 
 * The string passed in is assumed to be a class name - it is not checked.
 * If the class is unpackaged, return an empty string.
 * 
 * @param className the className to get the package name for, may be null
 * @return the package name or an empty string
 */
\end{lstlisting} 
            \subcaption{Javadoc}
\begin{lstlisting}[language=Java]
public void test()  throws Throwable  {
    String string0 = ClassUtils.getPackageName("line.separator");
    assertEquals("ine", string0);
}
\end{lstlisting} 
            \subcaption{Regression Test Case Generated by EvoSuite}
        \end{subfigure}
    \end{tabular}
    \caption{An example of a buggy source code along with its corresponding Javadoc and EvoSuite-generated regression test case}
    \label{fig:example}
\end{figure*}
\lstset{
    basicstyle=\ttfamily\footnotesize
}

\subsection{Capturing Program behavior using Regression Tests}
We employ automatically generated regression tests to produce a textual representation of the current program behavior. This textual representation serves as input for \name, enabling us to compare the program's captured behavior against its specifications. Regression test generation tools, such as EvoSuite~\cite{Evosuite}, are typically used to generate tests that help ensure that future updates do not inadvertently disrupt the existing functionality of the program. 

\Cref{fig:example} illustrates an example of a regression test generated for the \texttt{ClassUtils.getPackageName} method in the Apache Commons Lang project, which has an incorrect oracle. This test demonstrates the consequences of an artificially introduced bug in the method's return statement. When \texttt{ClassUtils.getPackageName("line.separator")} is invoked with the current version of the program, it erroneously returns \texttt{"ine"}. This output is in conflict with the method's documented expected behavior, which is to accurately return \texttt{"line"} as the package name for the input.

\subsection{Prompt Engineering based on Metamorphic Relations}

\begin{table}[t]
    \centering
    \caption{Oracle transformations based on MR}
    \scalebox{0.9}{
    \begin{tabular}{l|l}
    \toprule
    \textbf{Transformation} & {\textbf{Description}} \\ \midrule
    MR\_T2F          & Replacing \texttt{assertTrue} to \texttt{assertFalse}             \\ 
    MR\_F2T          & Replacing \texttt{assertFalse} to \texttt{assertTrue}             \\ 
    MR\_N2NN         & Replacing \texttt{assertNull} to \texttt{assertNotNull}            \\ 
    MR\_NN2N         & Replacing \texttt{assertNotNull} to \texttt{assertNull}           \\ 
    MR\_E2NE         & Replacing \texttt{assertEquals} to \texttt{assertNotEquals}       \\ 
    MR\_NE2E         & Replacing \texttt{assertNotEquals} to \texttt{assertEquals}       \\ 
    MR\_S2NS         & Replacing \texttt{assertSame} to \texttt{assertNotSame}            \\ 
    MR\_NS2S         & Replacing \texttt{assertNotSame} to \texttt{assertSame}            \\ \bottomrule
    \end{tabular}
    }
    \label{tab:transformation}
\end{table}

\name uses Metamorphic Relations (MRs) to improve the reliability of the LLM responses. These MRs are grounded in two core components: \textit{the input transformation} and \textit{the output relation}~\cite{segura2016survey}. 
The MRs employed in \name are defined as follows:
\begin{equation}
\begin{split}
R = \{&(a_{1}, a_{2}, \textit{Exec}(t,a_{1}), \textit{Exec}(t,a_{2})) \mid a_{1}=\neg a_{2} \\
&\rightarrow \textit{Exec}(t,a_{1}) = \lnot \textit{Exec}(t,a_{2})\}
\end{split}
\end{equation}
where $a_{1}$ and $a_{2}$ are assertion predicates that negate each other, the function \textit{Exec} returns the execution result (true if pass, false if fail) for test input \textit{t} against the given assertion. Note that the execution results for \textit{t} with $a_{1}$ should be different from that of $a_{2}$ in order to satisfy the output relation.
To generate a transformed test case, we apply a transformation as described in \Cref{tab:transformation} to the source test case generated at Step B. Through \textit{input transformation}, we negate the semantics of the original assertion. Subsequently, to verify whether \textit{the output relation} is met, we compare test outcomes to ensure that a pass in one execution directly corresponds to a fail in its counterpart, and vice versa. Notably, we filter out original assertions that do not lead to clear metamorphic relations, such as assertions that expect thrown exceptions (e.g., \texttt{assertThrows}).

\begin{figure}[t]
    \centering
    \includegraphics[width=0.9\linewidth]{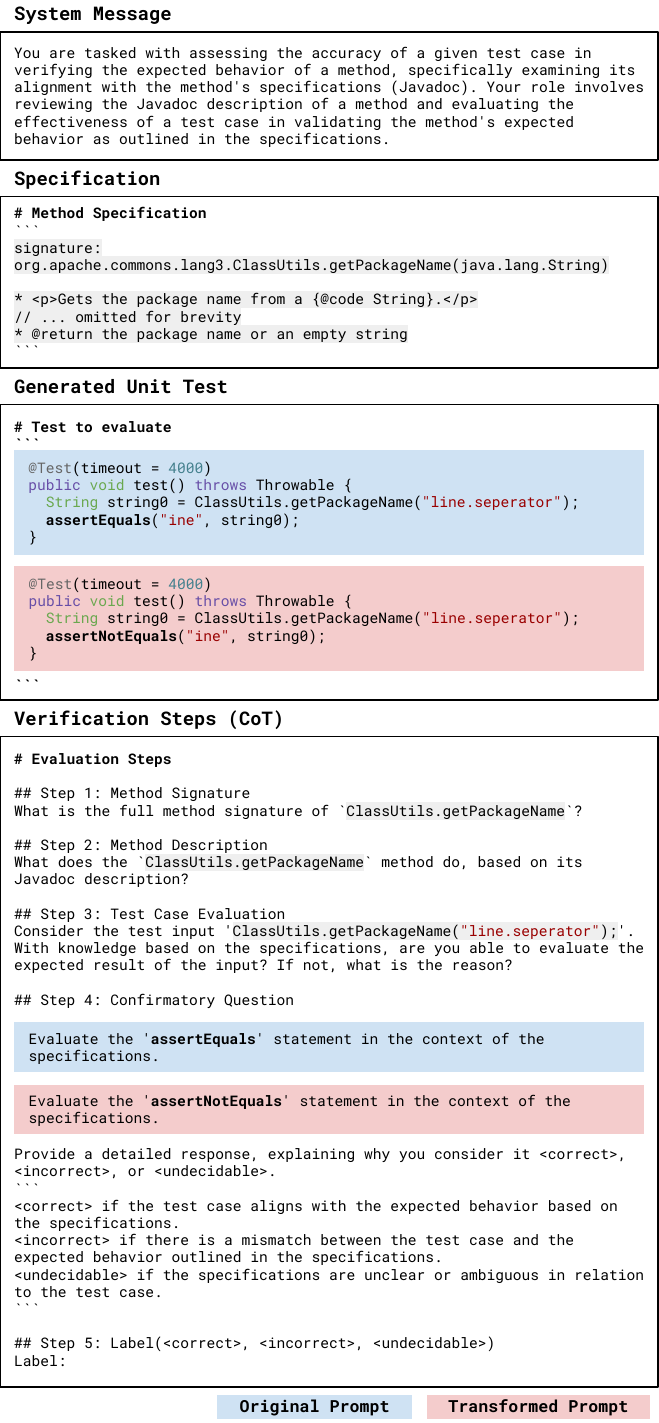}
    \caption{An example of metamorphic prompt}
    \label{fig:prompt}
\end{figure}

In the final stage of the prompt construction, we enhance the model's problem-solving capability by incorporating the Chain-of-Thought technique~\cite{Wei2024aa}. This approach explicitly outlines each reasoning step, thereby improving the model's analytical processes. It facilitates systematic evaluation of the model's reasoning and ensures clarity in its responses, significantly enhancing the reliability and interpretability of outcomes. The process involves several key steps:
\begin{itemize}
\item \textbf{Step 1:} Identifying Method Signature
\item \textbf{Step 2:} Identifying Method Description
\item \textbf{Step 3:} Evaluating Test Case
\item \textbf{Step 4:} Asking Confirmatory Question
\item \textbf{Step 5:} Labeling Oracle
\end{itemize}

We use the term \textit{metamorphic prompt} to refer to a pair of an original prompt and its transformed prompt. The prompt example referenced by \Cref{fig:example} is illustrated in \Cref{fig:prompt}. The texts with gray background are automatically filled in by \name for both the original and the transformed prompts, while the sections with blue and red backgrounds represent the content for the original and transformed prompts, respectively.

\subsection{Querying LLM}

Using the generated prompt, \name queries the LLM to respond to the reasoning steps (Step D). We present both the original version and the transformed-version prompt to the LLM, each version $n$ times, adopting the self-consistency prompt engineering. For the output, we instruct the LLM to label the correctness of the given test case and assertion with respect to the given documentation using three distinct types: $\texttt{<correct>}$, $\texttt{<undecidable>}$ and $\texttt{<incorrect>}$. Given that the test oracles (representing the program's behavior) in the original version and the transformed version are opposite to each other, we would expect the responses from an LLM to differ for each prompt if the LLM is consistent and can judge the consistency between the given document and the test case.

\subsection{Scoring based on LLM Responses}{\label{sec:methodology:scoring}}

We aggregate LLM responses and convert these responses into scores. The scoring methodology differs for original and transformed prompts, and is calculated as follows: 

\subsubsection{Original Prompts} For original prompts, the final label obtained from the LLM's responses is converted into a numerical form using the following function:
$$
f_{orig}(r) = \begin{cases}
+1, & \text{if } r = \texttt{<correct>}\\
0, & \text{if } r = \texttt{<undecidable>}\\
-1, & \text{if } r = \texttt{<incorrect>}
\end{cases}
$$
Suppose that we query the original-version prompt $n$ times and the answers are $\{r_1, \cdots, r_n\}$, we aggregate the answers by taking the sum of the numerical values, $\text{score}_{orig}(\{r_1, \cdots, r_n\}) = \sum_{i=1}^{n}f_{orig}(r)$, which ranges from $-n$ to $n$. This score represents the degree of alignment or misalignment between the program's actual behavior and its intended specification in documentation, as perceived by the LLM. A score of $n$ indicates that all $n$ responses from the LLM unanimously confirmed a consistency between the program behavior and specifications. Conversely, a score of $-n$ signifies that all responses identified an inconsistency.

\subsubsection{Transformed Prompts}
The scoring system for transformed prompts is deliberately inverted: 
$$
f_{tran}(r) = \begin{cases}
+1, & \text{if } r = \texttt{<incorrect>}\\
0, & \text{if } r = \texttt{<undecidable>}\\
-1, & \text{if } r = \texttt{<correct>}
\end{cases}
$$
Similarly, the answers from $n$ queries are aggregated as: $\text{score}_{tran}(\{r_1', \cdots, r_n'\}) = \sum_{i=1}^{n}f_{tran}(r')$, which also ranges from $-n$ to $n$, representing the degree of alignment or misalignment between the program's behavior and its specification. For instance, a score of $-n$ indicates that the LLM consistently found a consistency between the behavior and specification of the \textit{inversed} program across all $n$ responses. This, in turn, indicates an inconsistency between the behavior and specification of the original program, aligning with $f_{orig}$.

We then aggregate scores from both the original and transformed prompts by taking the sum of both scores. When querying both types of prompts $n$ times, the final score will range from $[-2n, 2n]$, where $2n$ represents a scenario where all $n$ responses for the original prompt are $\texttt{<correct>}$, and all $n$ responses for the transformed prompt are $\texttt{<incorrect>}$ (or vice versa). We then normalize this score to a score ranging in $[-1, +1]$ by dividing it by $2n$.


\begin{table}[h]
\centering
\caption{Details of \name dataset}
\resizebox{\columnwidth}{!}{
\begin{tabular}{l|r|rr|r}
\toprule
\multirow{2}{*}{\textbf{Projects}} & \multirow{2}{*}{\textbf{\# Mutants}} & \multicolumn{3}{c}{\textbf{\# Test}} \\
\cmidrule{3-5}
 & &\textbf{w/ incorrect oracle} & \textbf{w/ correct oracle} & \textbf{Total}\\\midrule
Chart         &   11,589    & 2,684      & 2,684         & 5,368           \\ 
Closure       &   343       & 93         & 93            & 186            \\ 
Lang          &   4,723     & 594        & 594           & 1,188          \\ 
Math          &   11,168    & 740        & 740           & 1,480          \\ 
Time          &   1,983     & 630        & 630           & 1,260          \\ \midrule
Total         &   29,806    & 4,741      & 4,741         & 9,482          \\ \bottomrule
\end{tabular}
}
\label{tab:dataset}
\end{table}

\section{Evaluation setup}
\label{sec:setup}

\subsection{Dataset}

We evaluate our approach on a carefully constructed dataset comprising $9,482$ pairs of tests and documentation. This dataset, shown in \Cref{tab:dataset}, is evenly balanced, containing $4,800$ tests with incorrect oracles and an equal number of tests with correct oracles. We construct this dataset from five open-source projects included in Defects4J v2.0.1 as follows:

\subsubsection{\textbf{Documentation Quality Assessment}} For each project, we examine the documentation quality of each method to confirm it contains descriptions for both parameters and return conditions within the latest fixed version for each project (e.g., Chart-1f, Closure-1f). Only methods with documentation satisfying these criteria move to the next step. As illustrated in \Cref{tab:dataset}, despite its large size, many methods from the Closure project were filtered out during this phase.

\subsubsection{\textbf{Mutant Injection}} 
This step artificially alters the program semantics to create inconsistencies between the code and its corresponding documentation, which are then used to evaluate our technique. Specifically, we generate first-order mutants using Major~\cite{just2011major}, a mutation testing tool for Java programs, in methods that have passed the documentation quality assessment. We aim to generate a broad range of inconsistent code-documentation pairs, as the original pairs from Defects4J do not cover sufficiently diverse inconsistencies.

\subsubsection{\textbf{Regression Test Generation}} Due to the high cost of generating regression test cases, we randomly choose no more than 10 mutants per method, prioritizing mutants that modify distinct lines of code to ensure diversity. However, if a method produces fewer than 10 mutants, we accept the set as is, without sampling. After selecting the mutants out of about 30,000 mutants from five projects, we employ EvoSuite to generate regression tests targeting the chosen mutated methods.

\subsubsection{\textbf{Oracle Identification}} Automatically generated tests are executed against the latest fixed version of the program, identifying the outcomes as failing or passing.
A failing test indicates a test with an incorrect oracle that captures the behavior modified by the mutants injected into the program, while a passing test has a correct oracle that conforms to the original program behavior.
Since some mutants may be difficult to kill, the number of failing tests is significantly smaller than that of the passing ones. To ensure a fair comparison, we randomly sample passing tests to match the number of failing tests.

\subsection{Experimental Settings}
As a LLM model, we use \texttt{GPT-3.5-Turbo-0613} provided by OpenAI with a default parameter setting of temperature 0.7.
We use EvoSuite version 1.0.7 and Major version 1.3.4 for test case and mutant generation, respectively.

\subsection{Research Questions}
We ask the following research questions in this paper.

\subsubsection{\textbf{RQ1. What is the effectiveness of \name?}} 

To answer this question, we apply \name (with $n=5$) to the 9,482 regression test cases shown in \Cref{tab:dataset}. Our analysis focuses on assessing how well the normalized score computed from each metamorphic prompt corresponds with the ground truth. We evaluate how effectively \name identifies misalignment between regression oracles and documentations by examining the precision and recall against different thresholds.

\subsubsection{\textbf{RQ2. How does each component affect the performance of \name?}} 
As described in \Cref{sec:methodology}, \name employs techniques such as metamorphic relations and self-consistency to enhance the reliability of the LLM. To examine the impact of these techniques on \name, we conduct an ablation study. This study also evaluates the essential role of the $\texttt{<undecidable>}$ label, which is used in cases where making a determination based solely on the provided prompt is difficult. We investigate the model's performance using only two labels, $\texttt{<correct>}$ and $\texttt{<incorrect>}$, to assess the utility of the $\texttt{<undecidable>}$ label.


\subsubsection{\textbf{RQ3. In what circumstance does \name fail to identify inconsistencies?}} 
In RQ3, we conduct a qualitative analysis to identify the environments in which \name fails to detect inconsistencies accurately. Specifically, we examine cases where \name reports consistencies with high confidence, even though the test-specification pairs were, in fact, inconsistent. This examination aims to uncover the causes behind these misjudgments.

\section{Results}
\label{sec:results}


\subsection{\textbf{RQ1.} Effectiveness of \name}

\begin{figure}[t]
    \centering 
    \includegraphics[width=\columnwidth]{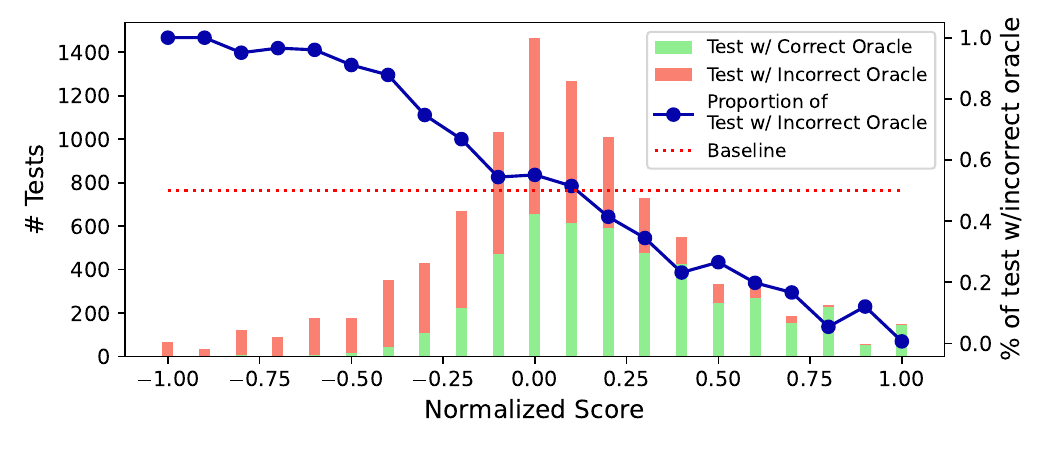}
    \caption{Incorrect oracle detection of \name}
    \label{fig:stack_bar}
\end{figure}

\begin{table}[t]
\centering
\caption{Precision, Recall, and F1 at different thresholds}
\scalebox{0.95}{
\begin{tabular}{lrrr|lrrr}
\toprule
\textbf{Score} & \textbf{Pre.} & \textbf{Rec.} & \textbf{F1} & \textbf{Score} & \textbf{Pre.} & \textbf{Rec.} & \textbf{F1} \\\midrule
$\leq -0.1$      & 0.722               & 0.480            & 0.576 & $\leq -0.6$      & 0.967               & 0.099            & 0.180     \\
$\leq -0.2$      & 0.808               & 0.361            & 0.499 & $\leq -0.7$      & 0.971               & 0.064            & 0.120     \\
$\leq -0.3$      & 0.873               & 0.267            & 0.409 & $\leq -0.8$      & 0.973               & 0.046            & 0.087     \\
$\leq -0.4$      & 0.926               & 0.199            & 0.328 & $\leq -0.9$      & 1.000               & 0.021            & 0.042     \\
$\leq -0.5$      & 0.952               & 0.134            & 0.235 & $\leq -1.0$      & 1.000               & 0.014            & 0.027      \\ \bottomrule
\end{tabular}}
\label{tab:metrics}
\end{table}

To answer RQ1, we present the results of \name (with $n=5$) evaluated on the five projects listed in \Cref{tab:dataset}. \Cref{fig:stack_bar} shows the distribution of the number of metamorphic prompts across the normalized score, with the red indicating those based on tests with incorrect oracles, and the green for correct ones. The majority of evaluations result in a score of 0.0, where the \name either responded $\texttt{<undecidable>}$ for all queries, or produced both positive and negative scores that canceled each other out. This shows the difficulty in assessing the correctness of oracles from developer-written documentation that might lack sufficient detail for clear evaluation (we used the documentation exactly as written by developers). The blue line depicts the proportion of metamorphic prompts constructed from tests with incorrect oracles (denoted as red in the bar plot) relative to all metamorphic prompts corresponding to each score. The red dotted line represents a baseline of 0.5, reflecting the balanced nature of our dataset, which includes an equal number of tests with correct and incorrect oracles. A point close to 100\% at the score -1.0 signifies that the test-specification pairs steadily identified by \name to be inconsistent were indeed pairs with inconsistencies. Conversely, the point near 0\% at the score 1.0 means that the pairs confidently classified by the \name as aligned are very rarely associated with an incorrect oracle.

To further evaluate the performance of \name in detecting the test oracles inconsistent with documentation, we analyze the precision and recall at different scoring thresholds, as detailed in \Cref{tab:metrics}. The oracle in a test is classified as \emph{incorrect} if the normalized score is equal to or lower than a specified threshold.
For example, when the threshold is set to -0.1, all instances with negative scores are classified as incorrect. At this threshold, the precision is 0.722 while the recall is 0.480. As the threshold is lowered, the precision increases but the recall drops.
These results indicate that choosing an appropriate threshold based on user requirements can balance utility and performance.



\begin{tcolorbox}[colback=gray!20, boxsep=0pt, left=1mm, right=1mm, top=1mm, bottom=1mm]
\textbf{Answer to RQ1:} \name effectively identified misalignment between documentation and test, demonstrating a high precision of 0.722 and a recall of 0.480 in detecting inconsistencies. When applying stricter thresholds, precision can be set to nearly 100\%.
\end{tcolorbox}

\subsection{\textbf{RQ2:} Ablation Study}

\begin{figure*}[t]
     \centering
     \begin{subfigure}[h]{0.30\linewidth}
         \centering
         \includegraphics[width=0.9\columnwidth]{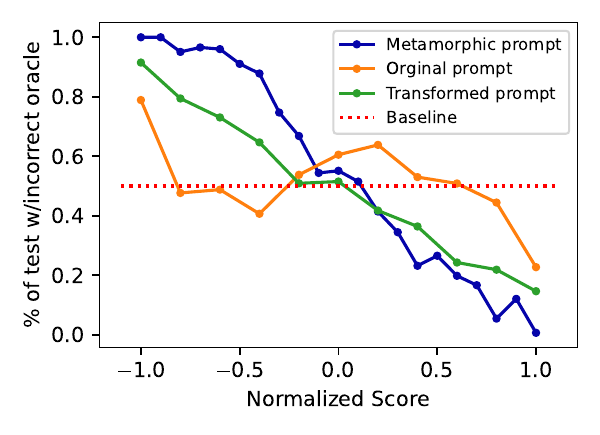}
         \caption{metamorphic}
         \label{fig:metamorphic_ratio}
     \end{subfigure}
     \begin{subfigure}[h]{0.30\linewidth}
         \centering
         \includegraphics[width=0.9\columnwidth]{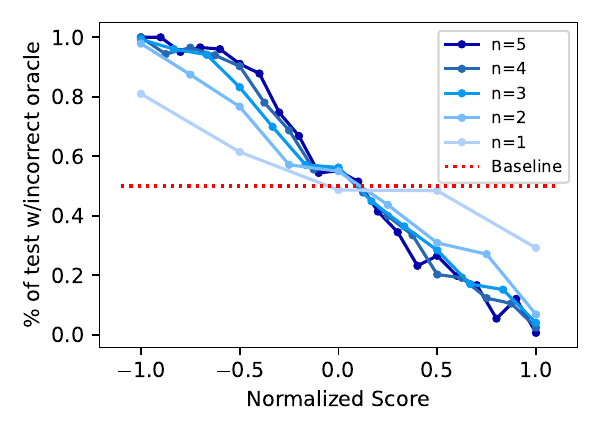}
         \caption{self-consistency}
         \label{fig:selfconsistency}
     \end{subfigure}
     \begin{subfigure}[h]{0.30\linewidth}
         \centering
         \includegraphics[width=0.9\columnwidth]{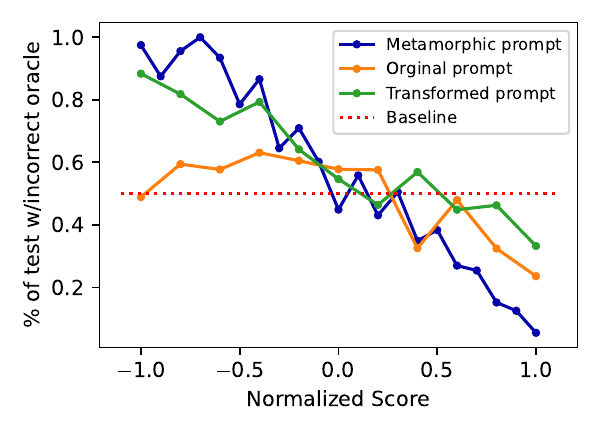}
         \caption{without $\texttt{<undecidable>}$ label}
         \label{fig:metamorphic_ratio_two}
     \end{subfigure}
        \caption{An impact of metamorphic relations, self-consistency, and labels}
        \label{fig:metamorphic_selfconsistency}
\end{figure*}

\Cref{fig:metamorphic_ratio} shows the ratio of tests with incorrect oracles per their scores when \name is applied using only the original (orange) and the transformed (green) prompts. The blue line and red-dotted line follow the same representation as in RQ1. If metamorphic relations are not employed, i.e., assessing inconsistencies with just original prompts, the capability of \name to detect incorrect oracles shows a marked decrease. Specifically, when scores range between -1.0 and 0.0, the ratio drops below the baseline, leading to outcomes that are essentially indistinguishable from random alignment judgments. However, when incorporating the outcomes from the transformed prompts, as observed in RQ1, it is almost always the case that prompts receiving low scores from \name are strongly associated with incorrect oracles. 

\Cref{fig:selfconsistency} illustrates the impact of the number of queries to the LLM on the performance of \name. When the number of queries, denoted as \textit{n} in \Cref{sec:methodology:scoring}, for both the original and transformed prompts increases, we observe a higher ratio of incorrect oracles within the score ranges from -1.0 to 0.0, and a lower ratio of incorrect oracles within the range of [0,1]. Additionally, the observed gain in performance grows smaller as $n$ increases, suggesting that \name may be converging with respect to the number of queries. 



\begin{table}[t]
    \centering
    \caption{Spearman's correlation coefficient ($\rho$) and the p-value between scores and the ratio of incorrect oracles}
        \scalebox{0.95}{
        \begin{tabular}{l|rr|rr}
        \toprule
          & \multicolumn{2}{c|}{w/o $\texttt{<undecidable>}$} & \multicolumn{2}{c}{w/ $\texttt{<undecidable>}$}\\\cline{2-5}
          & $\mathbf{\rho}$ & p-value & $\mathbf{\rho}$ & p-value \\ 
        \midrule
        Metamorphic Prompt & -0.977  & 3.81e-13 & \textbf{-0.992}  & 1.75e-18\\
        Original Prompt    & -0.700  & 1.65e-02 & -0.355  & 2.84e-01\\
        Transformed Prompt & -0.955  & 4.99e-06 & -0.991  & 3.76e-09 \\
        \bottomrule
        \end{tabular}
        }
    \label{tab:correlation}
\end{table}

We also explore the need for the $\texttt{<undecidable>}$ label. \Cref{fig:metamorphic_ratio_two} shows the results obtained using \name but without the $\texttt{<undecidable>}$ label, i.e., the LLM is forced to label each assertion as $\texttt{<correct>}$ or $\texttt{<incorrect>}$. It shows a weaker correlation between the scores and the ratio of oracles classified as incorrect at each score. This observation is supported by the Spearman's correlation coefficient values presented in \Cref{tab:correlation}. Spearman's correlation coefficient measures the strength and direction of a monotonic relationship between two variables. As can be seen in \Cref{tab:correlation}, including $\texttt{<undecidable>}$ produces a stronger negative correlation across metamorphic prompts as evidenced by Spearman's correlation coefficient values approaching -1.
These findings suggest that including the $\texttt{<undecidable>}$ label can improve the effectiveness of our approach, especially in situations where the program specifications in the documentation may not clearly define the program's semantics.
\begin{tcolorbox}[colback=gray!20, boxsep=0pt, left=1mm, right=1mm, top=1mm, bottom=1mm]
\textbf{Answer to RQ2:} The ablation study shows that metamorphic relations, self-consistency, and the $\texttt{<undecidable>}$ label enhance the effectiveness of \name.
\end{tcolorbox}

\begin{table}[t]
    \centering
    \caption{Analysis of false alarms in \name}
    \label{tab:llm_performance}
    \scalebox{0.9}{
    \begin{tabular}{lcccc|c}
    \toprule
     & \textbf{Chart} & \textbf{Lang} & \textbf{Math} & \textbf{Time} & \textbf{Total}\\
    \midrule
    Lack of Specification & 2 & 0 & 0 & 0 & 2\\
    Need for Contextual information    & 1 & 0 & 2 & 0 & 3\\ 
    LLM Underperformance  & 0 & 8 & 7 & 1 & 16\\
    \bottomrule
    \end{tabular}
    }
\end{table}

\begin{figure}[t]
    \centering
\begin{lstlisting}[basicstyle=\ttfamily\scriptsize, language=Java]
/*
 * Returns a string that is equivalent to the input string,
 * but with special characters converted to JavaScript
 * escape sequences. < ... Omitted ...>
 */

\end{lstlisting} 
\begin{lstlisting}[basicstyle=\ttfamily\scriptsize, language=Java]
public void test()  throws Throwable  {
  String string0 = ImageMapUtilities.javascriptEscape("&lt;");
  assertEquals("\\'lt;", string0);
}
\end{lstlisting} 
    \caption{An example of \textit{Lack of Specification Detail}}
    \label{fig:specification}
\end{figure}

\subsection{\textbf{RQ3:} Qualitative Analysis}

In RQ3, we analyze instances where \name identified tests with incorrect oracles as correct, reflected by normalized scores of 0.8 or higher, denoting high confidence. Among the 9,482 pairs analyzed, 21 were identified as exhibiting this discrepancy, and we manually inspected the reasons behind these. We categorize the causes of misclassification into three primary reasons, \textit{Lack of Specification Detail}, \textit{Need for Contextual Information}, and \textit{LLM Underperformance}, whose distributions across projects are shown in \Cref{tab:llm_performance}.

\begin{figure}[t]
    \centering
\begin{lstlisting}[basicstyle=\ttfamily\scriptsize, language=Java]
/*signature: org.apache.commons.math3.geometry.euclidean.threed.Plane.isSimilarTo

 * Check if the instance is similar to another plane.
 * <p>Planes are considered similar if they contain the same
 * points. This does not mean they are equal since they can have
 * opposite normals.</p>
 * @param plane plane to which the instance is compared
 * @return true if the planes are similar
*/
\end{lstlisting} 
\begin{lstlisting}[basicstyle=\ttfamily\scriptsize, language=Java]
public void test3()  throws Throwable  {
  Vector3D vector3D0 = Vector3D.NaN;
  Plane plane0 = new Plane(vector3D0);
  boolean boolean0 = plane0.isSimilarTo(plane0);
  assertTrue(boolean0);
}
\end{lstlisting} 
    \caption{An example of \textit{Need for Contextual Information}}
    \label{fig:complexity}
\end{figure}

\noindent\textbf{Lack of Specification Detail:} An example of this category is in \Cref{fig:specification}. Although the specification mentions that the method is supposed to convert special characters to JavaScript escape sequences, it does not provide concrete examples of special characters. This lack of detailed specification makes it challenging to assess the correctness of the test oracles.

\begin{figure}[t]
\begin{lstlisting}[basicstyle=\ttfamily\scriptsize, language=Java]
public void test()  throws Throwable  {
  Fraction fraction0 = Fraction.ONE_HALF;
  Fraction fraction1 = Fraction.ONE_THIRD;
  int int0 = fraction0.compareTo(fraction1);
  assertEquals((-1), int0);
}
\end{lstlisting}
    \includegraphics[width=\columnwidth]{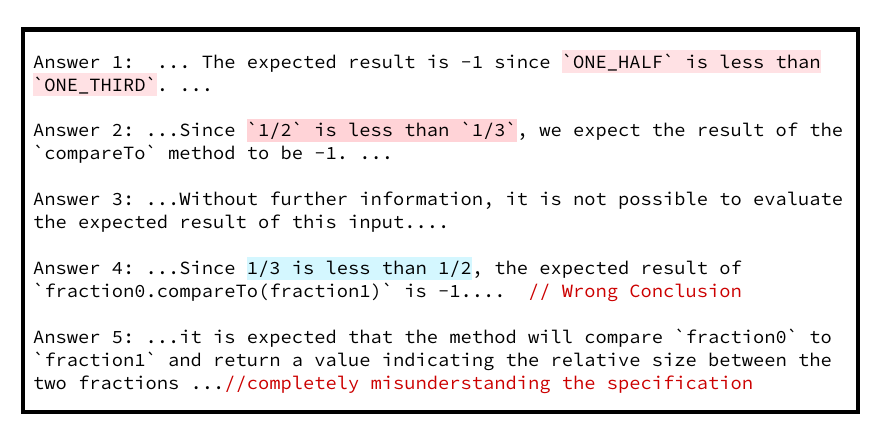}
    \caption{An example of \textit{Underperformace of LLMs}}
    \label{fig:underperformance}
\end{figure}

\noindent\textbf{Need for Contextual Information:} An example of this category is in \Cref{fig:complexity} containing the specification of the \texttt{isSimilarTo} function from \texttt{Math}. As the properties that a Plan object acquires upon creation are not known, it becomes challenging to accurately evaluate the test outcomes. 

\noindent\textbf{LLM Underperformance:} \Cref{fig:underperformance} shows a simple test case that the LLM fails to classify correctly due to its lack of mathematical reasoning capabilities. However, the reasoning abilities of LLMs have been improving over time. When we re-evaluated all prompts associated with this kind of case using GPT-4, the average normalized score improved to $-0.156$, reflecting the improved arithmetic capabilities, suggesting that such issues may further diminish as LLM performance continues to enhance.

\begin{tcolorbox}[colback=gray!20, boxsep=0pt, left=1mm, right=1mm, top=1mm, bottom=1mm]
\textbf{Answer to RQ3:} The majority of false alarms were due to the limitations in the reasoning capabilities of the LLMs. Additionally, depending on projects, factors such as the lack of detailed specifications and the need for contextual information also contributed to these issues.
\end{tcolorbox}





\section{Conclusion}
\label{sec:conclusion}
This paper introduces \name, an LLM-based technique designed to automatically identify inconsistencies between a program's documentation and its actual behavior, as captured by regression test oracles. To address the issue of LLM hallucinations, \name captures program behavior via the regression test cases generated by EvoSuite and applies metamorphic prompts. Our experiments with 9,482 test-documentation pairs derived from Defects4J show that \name can effectively identify inconsistency with a precision of 0.72 and a recall of 0.48. We hope to expand upon these results by exploring its capabilities when used in conjunction with existing techniques such as fault localization and automatic program repair.

\bibliographystyle{IEEEtran}
\bibliography{ref}

\begin{thebibliography}{10}
\providecommand{\url}[1]{#1}
\csname url@samestyle\endcsname
\providecommand{\newblock}{\relax}
\providecommand{\bibinfo}[2]{#2}
\providecommand{\BIBentrySTDinterwordspacing}{\spaceskip=0pt\relax}
\providecommand{\BIBentryALTinterwordstretchfactor}{4}
\providecommand{\BIBentryALTinterwordspacing}{\spaceskip=\fontdimen2\font plus
\BIBentryALTinterwordstretchfactor\fontdimen3\font minus \fontdimen4\font\relax}
\providecommand{\BIBforeignlanguage}[2]{{%
\expandafter\ifx\csname l@#1\endcsname\relax
\typeout{** WARNING: IEEEtran.bst: No hyphenation pattern has been}%
\typeout{** loaded for the language `#1'. Using the pattern for}%
\typeout{** the default language instead.}%
\else
\language=\csname l@#1\endcsname
\fi
#2}}
\providecommand{\BIBdecl}{\relax}
\BIBdecl

\bibitem{de2006documentation}
S.~C.~B. de~Souza, N.~Anquetil, and K.~M. de~Oliveira, ``Which documentation for software maintenance?'' \emph{Journal of the Brazilian Computer Society}, vol.~12, pp. 31--44, 2006.

\bibitem{evolutionOfCodeComments}
\BIBentryALTinterwordspacing
Z.~M. Jiang and A.~E. Hassan, ``Examining the evolution of code comments in postgresql,'' in \emph{Proceedings of the 2006 International Workshop on Mining Software Repositories}, ser. MSR '06.\hskip 1em plus 0.5em minus 0.4em\relax New York, NY, USA: Association for Computing Machinery, 2006, p. 179–180. [Online]. Available: \url{https://doi.org/10.1145/1137983.1138030}
\BIBentrySTDinterwordspacing

\bibitem{tan2007icomment}
L.~Tan, D.~Yuan, G.~Krishna, and Y.~Zhou, ``/* icomment: Bugs or bad comments?*,'' in \emph{Proceedings of twenty-first ACM SIGOPS symposium on Operating systems principles}, 2007, pp. 145--158.

\bibitem{tan2011acomment}
L.~Tan, Y.~Zhou, and Y.~Padioleau, ``acomment: mining annotations from comments and code to detect interrupt related concurrency bugs,'' in \emph{Proceedings of the 33rd international conference on software engineering}, 2011, pp. 11--20.

\bibitem{tan2012tcomment}
S.~H. Tan, D.~Marinov, L.~Tan, and G.~T. Leavens, ``@ tcomment: Testing javadoc comments to detect comment-code inconsistencies,'' in \emph{2012 IEEE Fifth International Conference on Software Testing, Verification and Validation}.\hskip 1em plus 0.5em minus 0.4em\relax IEEE, 2012, pp. 260--269.

\bibitem{ratol2017detecting}
I.~K. Ratol and M.~P. Robillard, ``Detecting fragile comments,'' in \emph{2017 32nd IEEE/ACM International Conference on Automated Software Engineering (ASE)}.\hskip 1em plus 0.5em minus 0.4em\relax IEEE, 2017, pp. 112--122.

\bibitem{corazza2018coherence}
A.~Corazza, V.~Maggio, and G.~Scanniello, ``Coherence of comments and method implementations: a dataset and an empirical investigation,'' \emph{Software Quality Journal}, vol.~26, pp. 751--777, 2018.

\bibitem{panthaplackel2021deep}
S.~Panthaplackel, J.~J. Li, M.~Gligoric, and R.~J. Mooney, ``Deep just-in-time inconsistency detection between comments and source code,'' in \emph{Proceedings of the AAAI Conference on Artificial Intelligence}, vol.~35, no.~1, 2021, pp. 427--435.

\bibitem{rabbi2020detecting}
F.~Rabbi and M.~S. Siddik, ``Detecting code comment inconsistency using siamese recurrent network,'' in \emph{Proceedings of the 28th International Conference on Program Comprehension}, 2020, pp. 371--375.

\bibitem{li2022competition}
Y.~Li, D.~Choi, J.~Chung, N.~Kushman, J.~Schrittwieser, R.~Leblond, T.~Eccles, J.~Keeling, F.~Gimeno, A.~Dal~Lago \emph{et~al.}, ``Competition-level code generation with alphacode,'' \emph{Science}, vol. 378, no. 6624, pp. 1092--1097, 2022.

\bibitem{kang2023large}
S.~Kang, J.~Yoon, and S.~Yoo, ``Large language models are few-shot testers: Exploring llm-based general bug reproduction,'' in \emph{2023 IEEE/ACM 45th International Conference on Software Engineering (ICSE)}.\hskip 1em plus 0.5em minus 0.4em\relax IEEE, 2023, pp. 2312--2323.

\bibitem{schafer2023adaptive}
M.~Sch{\"a}fer, S.~Nadi, A.~Eghbali, and F.~Tip, ``Adaptive test generation using a large language model,'' \emph{arXiv e-prints}, pp. arXiv--2302, 2023.

\bibitem{li2024mutation}
Z.~Li and D.~Shin, ``Mutation-based consistency testing for evaluating the code understanding capability of llms,'' \emph{arXiv preprint arXiv:2401.05940}, 2024.

\bibitem{huang2023survey}
L.~Huang, W.~Yu, W.~Ma, W.~Zhong, Z.~Feng, H.~Wang, Q.~Chen, W.~Peng, X.~Feng, B.~Qin \emph{et~al.}, ``A survey on hallucination in large language models: Principles, taxonomy, challenges, and open questions,'' \emph{arXiv preprint arXiv:2311.05232}, 2023.

\bibitem{Evosuite}
\BIBentryALTinterwordspacing
G.~Fraser and A.~Arcuri, ``Evosuite: automatic test suite generation for object-oriented software,'' in \emph{Proceedings of the 19th ACM SIGSOFT Symposium and the 13th European Conference on Foundations of Software Engineering}, ser. ESEC/FSE '11.\hskip 1em plus 0.5em minus 0.4em\relax New York, NY, USA: Association for Computing Machinery, 2011, p. 416–419. [Online]. Available: \url{https://doi.org/10.1145/2025113.2025179}
\BIBentrySTDinterwordspacing

\bibitem{wei2022chain}
J.~Wei, X.~Wang, D.~Schuurmans, M.~Bosma, F.~Xia, E.~Chi, Q.~V. Le, D.~Zhou \emph{et~al.}, ``Chain-of-thought prompting elicits reasoning in large language models,'' \emph{Advances in neural information processing systems}, vol.~35, pp. 24\,824--24\,837, 2022.

\bibitem{wang2022self}
X.~Wang, J.~Wei, D.~Schuurmans, Q.~Le, E.~Chi, S.~Narang, A.~Chowdhery, and D.~Zhou, ``Self-consistency improves chain of thought reasoning in language models,'' \emph{arXiv preprint arXiv:2203.11171}, 2022.

\bibitem{just2014defects4j}
R.~Just, D.~Jalali, and M.~D. Ernst, ``Defects4j: A database of existing faults to enable controlled testing studies for java programs,'' in \emph{Proceedings of the 2014 international symposium on software testing and analysis}, 2014, pp. 437--440.

\bibitem{shin2023prompt}
J.~Shin, C.~Tang, T.~Mohati, M.~Nayebi, S.~Wang, and H.~Hemmati, ``Prompt engineering or fine tuning: An empirical assessment of large language models in automated software engineering tasks,'' \emph{arXiv preprint arXiv:2310.10508}, 2023.

\bibitem{sun2023automatic}
W.~Sun, C.~Fang, Y.~You, Y.~Miao, Y.~Liu, Y.~Li, G.~Deng, S.~Huang, Y.~Chen, Q.~Zhang \emph{et~al.}, ``Automatic code summarization via chatgpt: How far are we?'' \emph{arXiv preprint arXiv:2305.12865}, 2023.

\bibitem{ahmed2023improving}
T.~Ahmed, K.~S. Pai, P.~Devanbu, and E.~T. Barr, ``Improving few-shot prompts with relevant static analysis products,'' \emph{arXiv preprint arXiv:2304.06815}, 2023.

\bibitem{achiam2023gpt}
J.~Achiam, S.~Adler, S.~Agarwal, L.~Ahmad, I.~Akkaya, F.~L. Aleman, D.~Almeida, J.~Altenschmidt, S.~Altman, S.~Anadkat \emph{et~al.}, ``Gpt-4 technical report,'' \emph{arXiv preprint arXiv:2303.08774}, 2023.

\bibitem{segura2016survey}
S.~Segura, G.~Fraser, A.~B. Sanchez, and A.~Ruiz-Cort{\'e}s, ``A survey on metamorphic testing,'' \emph{IEEE Transactions on software engineering}, vol.~42, no.~9, pp. 805--824, 2016.

\bibitem{applis2021assessing}
L.~Applis, A.~Panichella, and A.~van Deursen, ``Assessing robustness of ml-based program analysis tools using metamorphic program transformations,'' in \emph{2021 36th IEEE/ACM International Conference on Automated Software Engineering (ASE)}.\hskip 1em plus 0.5em minus 0.4em\relax IEEE, 2021, pp. 1377--1381.

\bibitem{murphy2008properties}
C.~Murphy, G.~E. Kaiser, and L.~Hu, ``Properties of machine learning applications for use in metamorphic testing,'' 2008.

\bibitem{xie2011testing}
X.~Xie, J.~W. Ho, C.~Murphy, G.~Kaiser, B.~Xu, and T.~Y. Chen, ``Testing and validating machine learning classifiers by metamorphic testing,'' \emph{Journal of Systems and Software}, vol.~84, no.~4, pp. 544--558, 2011.

\bibitem{Weyuker1982aa}
E.~J. Weyuker, ``{On Testing Non-Testable Programs},'' \emph{The Computer Journal}, vol.~25, no.~4, pp. 465--470, 11 1982.

\bibitem{Wei2024aa}
J.~Wei, X.~Wang, D.~Schuurmans, M.~Bosma, B.~Ichter, F.~Xia, E.~H. Chi, Q.~V. Le, and D.~Zhou, ``Chain-of-thought prompting elicits reasoning in large language models,'' in \emph{Proceedings of the 36th International Conference on Neural Information Processing Systems}, ser. NIPS '22.\hskip 1em plus 0.5em minus 0.4em\relax Red Hook, NY, USA: Curran Associates Inc., 2024.

\bibitem{just2011major}
R.~Just, F.~Schweiggert, and G.~M. Kapfhammer, ``Major: An efficient and extensible tool for mutation analysis in a java compiler,'' in \emph{2011 26th IEEE/ACM International Conference on Automated Software Engineering (ASE 2011)}.\hskip 1em plus 0.5em minus 0.4em\relax IEEE, 2011, pp. 612--615.

\end{thebibliography}
\end{document}